\documentclass[twocolumn,10pt]{article}

\usepackage[margin=1.9cm,top=2.2cm,bottom=2.5cm]{geometry}
\usepackage{amsmath,amssymb,amsfonts}
\usepackage{graphicx}
\usepackage{booktabs}
\usepackage{multirow}
\usepackage{xcolor}
\usepackage{hyperref}
\usepackage{cleveref}
\usepackage{microtype}
\usepackage{caption}
\usepackage{subcaption}
\usepackage{float}
\usepackage{authblk}
\usepackage{makecell}

\hypersetup{
	colorlinks=true,
	citecolor=blue!70!black,
	linkcolor=blue!70!black,
	urlcolor=blue!70!black
}

\newcommand{\dG}{$\Delta G_{\mathrm{H}^*}$}

\newcommand{\HER}{hydrogen evolution reaction}

\begin{document}
	
	\title{\textbf{Generative Design and Graph Neural Network
			Screening of Novel Bimetallic Alloy Catalysts for the
			Hydrogen Evolution Reaction}}
	
	\author[1]{Vipin Kurian Elavunkel}
	\author[2]{Sylvain Le Tonquesse}
	\author[1,$\dagger$]{Hadrien Vroylandt}
	\affil[1]{Universit\'{e} Caen Normandie, ENSICAEN, CNRS,
		Normandie Univ, GREYC UMR 6072, F-14000 Caen, France}
	\affil[2]{Laboratoire CRISMAT, CNRS UMR 6508, ENSICAEN,
		Normandie Universit\'{e}, Caen 14050, France}
	\affil[$\dagger$]{\texttt{hadrien.vroylandt@unicaen.fr}}
	
	\date{\today}
	\maketitle
	
	\begin{abstract}
		\noindent
		Identifying earth-abundant catalysts for the \HER{} (HER) is a
		central challenge in scalable green hydrogen production.
		We present an end-to-end computational pipeline that couples
		diffusion-based generative modelling, machine-learning
		force-field relaxation, and graph neural network (GNN) property
		prediction to discover bimetallic HER catalysts beyond
		the structures and stoichiometries covered by existing adsorption databases.
		A generative model (MatterGen) is fine-tuned on bulk alloy
		structures labelled with their minimum hydrogen adsorption
		energy and steered towards adsorption energies near zero. From the
		generated pool we retain the stable structures that are
		novel bimetallic alloys: binary compositions built from the
		training-set metals but at stoichiometries or structures absent
		from that database. Each candidate's low-index facets are ranked by a
		machine-learning-force-field (MatterSim) surface-energy screen, and
		only the most stable terminations are converted into explicit slab
		models with hydrogen adsorbed at all high-symmetry surface sites and
		relaxed with the same force field. Two GNN surrogates
		(MEGNet and ALIGNN), trained on a high-throughput density functional theory (DFT)
		dataset,
		then screen the resulting slab candidates. Because the two models encode
		the local environment through different inductive biases, we retain
		only the alloys on which both agree, reducing the influence of any single model. This
		consensus screening, combined with an additional MatterSim
		energy-difference triangulation, resulted in a shortlist of
		near-thermoneutral bimetallic candidates. Eight are validated with
		plane-wave DFT, and four of them, Pd$_5$Au, Pt$_5$Rh, Cu$_5$Pt$_3$, and
		Pd$_{13}$Ru$_3$, bind hydrogen within the thermoneutral window. All eight alloys are dynamically stable by phonon calculations.
		These four alloys are highlighted as DFT-validated bimetallic HER
		candidates for experimental follow-up.
	\end{abstract}
	
	
	\section{Introduction}
	\label{sec:intro}
	
	The global transition to carbon-neutral energy demands scalable,
	cost-effective routes to green hydrogen. Electrochemical water
	splitting driven by renewable electricity is widely regarded as
	the most promising method, yet its commercial viability depends on the performance of the cathode catalyst for the
	\HER{} (HER)~\cite{norskov2005trends,greeley2006computational}. Pure
	platinum remains the benchmark HER material, but its scarcity
	and cost make large-scale
	deployment impractical~\cite{feidenhans2024pmfree}. The search for
	earth-abundant alternatives has therefore become one of the
	central challenges in
	electrocatalysis~\cite{kumar2024focused,deshmukh2025bimetallic},
	including transition metal dichalcogenides, carbides, phosphides,
	single-atom catalysts, and bimetallic
	alloys~\cite{liang2025sac,feidenhans2024pmfree}. Among these
	platforms, bimetallic alloys stand out for their ability to
	combine two distinct chemical functions in a single material:
	one metal may optimise hydrogen adsorption while the other
	promotes water dissociation, hydroxyl binding, or structural
	stability~\cite{kumar2024focused,sinha2024mlher}.
	
	The theoretical foundation for bimetallic HER design rests on
	the $d$-band centre model: the average energy of the transition
	metal $d$-band relative to the Fermi level correlates directly
	with adsorbate binding strength through the Sabatier principle,
	with maximum catalytic activity achieved at an intermediate
	$d$-band
	position~\cite{hammer1995gold,escudero2016tuning}.
	Alloying two metals modifies this centre through two cooperative
	mechanisms. The ligand (charge-transfer) effect arises
	from electronegativity differences between the two metals: the
	more electropositive component donates electron density to its
	partner, shifting the acceptor's $d$-band and weakening its
	adsorbate
	binding~\cite{wang2018strain,escudero2016tuning}.
	The strain effect arises from lattice mismatch: tensile
	strain narrows the $d$-band and raises its centre,
	strengthening binding, while compressive strain broadens it and
	lowers the centre, weakening
	binding~\cite{wang2018strain}. Pd monolayers deposited on different
	single-crystal substrates demonstrate this directly: their hydrogen
	evolution rate varies systematically with the lateral strain imposed by
	the substrate and the resulting shift of the Pd $d$-band
	centre~\cite{kibler2005tuning}. A recent example of these
	principles is provided by porous NiW bimetallic alloys
	fabricated via dynamic H$_2$ bubble templating: alloying Ni
	with W shifts the W~5$d$ band centre downward, simultaneously
	enhancing OH$^*$ desorption at W sites and H$_2$O adsorption
	and dissociation, while Ni$_2$W$_1$ hollow sites serve as the
	primary hydrogen production centres~\cite{yu2025niw}. This
	dual-site mechanism facilitates industrial-level current densities
	of 500 and 1{,}000~mA~cm$^{-2}$ at overpotentials of only 198
	and 264~mV respectively in 1.0~M KOH, with stable operation over a
	200-hour test~\cite{yu2025niw}. Beyond these electronic effects,
	the geometric ensemble effect governs which reaction
	intermediates can bind cooperatively at sites with specific arrangements of the two metal species, providing an
	additional handle for activity
	tuning~\cite{kumar2024focused,deshmukh2025bimetallic}. High-entropy
	alloys, which combine five or more elements in roughly
	equimolar proportions, extend this design space further by
	providing high configurational entropy that stabilises
	single-phase solid solutions and creates broad active-site diversity~\cite{chen2025hea}.
	
	These combined processes can push the hydrogen adsorption
	free energy \dG{} toward the thermoneutral ideal, enabling
	non-precious-metal-rich systems to approach Pt activity. The Ni--Mo
	alloy MoNi$_4$, for example, reaches an activity comparable to Pt in
	alkaline electrolyte~\cite{zhang2017moni4}, and earth-abundant
	bimetallic electrocatalysts have demonstrated competitive
	performance for overall water splitting in both acidic and alkaline
	media~\cite{zang2023pyrolysis}. Ordered
	intermetallic compounds, where each metal occupies a specific
	crystallographic site, offer a further advantage: well-defined
	metal-metal bonding confers superior resistance to
	dealloying, the selective dissolution of the less noble
	component that remains the principal stability challenge
	restricting on-site implementation of bimetallic
	catalysts~\cite{kumar2024focused,zhao2024engineering}.
	Mitigation strategies for dealloying, including controlled
	post-synthesis etching, protective surface coatings, and
	dopant stabilisation, represent active and as-yet unresolved
	areas of
	research~\cite{kumar2024focused,deshmukh2025bimetallic}.
	
	Despite these advances, the compositional and structural space
	of bimetallic alloys is enormous. For a binary transition-metal
	system, performance varies continuously across all compositions
	and is further modulated by crystal structure (random alloy,
	ordered intermetallic, core--shell nanostructure), surface facet, and
	nanoscale morphology, easily yielding hundreds of candidates
	per system to synthesise and
	test~\cite{kumar2024focused}. Traditional high-throughput density functional theory (DFT)
	screening, exemplified by the Mamun dataset~\cite{mamun2019high}
	that underpins this work, has mapped adsorption energetics
	across 37 metals and their bimetallic combinations,
	but is restricted to a small set of prototype structures
	(pure metals and A$_3$B and AB ordered alloys). The far larger space of novel, as-yet-unknown
	alloys remains largely unexplored.
	
	Machine learning is now a primary tool for accelerating the discovery process. Machine-learning models trained on DFT
	datasets can identify complex structure--property relationships
	and perform high-throughput screening of large compositional
	spaces at computational speeds orders of magnitude faster than
	DFT~\cite{sinha2024mlher,recent_advances_ml_alloy_her}.
	Recent work demonstrates this capability across several
	directions. Oguz~et~al.~\cite{oguz2025ml} integrated the
	equivariant neural network EquiformerV2 into an AdsorbML
	surrogate pipeline, screening thousands of bimetallic surfaces
	and identifying four synthesis-ready earth-abundant
	candidates (Fe$_2$Sb$_4$, Cu$_6$Sb$_2$, Cu$_6$Sn$_2$, and
	Ni$_2$Sb$_2$) with platinum-like activity and a validated mean absolute error (MAE)
	of 0.12~eV against DFT references. At the level of individual
	descriptors, an extremely randomized trees model using only ten
	features, including an energy-related descriptor built from the
	$d$-orbital occupancy, has predicted HER activity across diverse
	catalyst types with a coefficient of determination $R^2$ of 0.922,
	at about 1/200{,}000 of the cost of
	DFT~\cite{npj2025mlminfeature}. Machine-learning-assisted workflows have
	also been applied to single-atom catalyst design, where a Bayesian
	genetic-algorithm model identified Ni-incorporated carbon quantum dots with
	experimentally validated overpotentials of 151~mV at
	10~mA~cm$^{-2}$~\cite{carbonenergy2025mlsac}. Reviews of
	machine-learning-driven alloy catalyst discovery confirm that this approach has now been validated across binary, ternary, and high-entropy
	alloy systems, with models capable of uncovering non-obvious
	optimal compositions that would be inaccessible through
	conventional screening~\cite{recent_advances_ml_alloy_her,sinha2024mlher}.
	
	Two further advances open a route to entirely novel structural
	space beyond what DFT databases contain. First,
	diffusion-based generative models for inorganic crystals have advanced to the point of being usable for targeted materials discovery:
	MatterGen~\cite{zeni2025generative} jointly generates atomic
	species, fractional coordinates, and periodic lattice
	parameters through a denoising diffusion process conditioned on
	arbitrary scalar properties, enabling inverse design of
	structures with prescribed catalytic potential. Second, graph
	neural networks (GNNs) have become a widely used surrogate model for
	rapid DFT-quality property prediction on crystalline
	materials~\cite{xie2018crystal,chen2019graph,choudhary2021atomistic}.
	GNNs achieve this by automatically learning topological and
	geometric features from atomic graphs without manual feature
	engineering, and have been considered as a powerful approach for electrocatalyst design, facilitating accurate
	prediction of adsorption energies, reaction barriers, and
	stability metrics across diverse catalyst
	families~\cite{chen2025gnn,jiao2024gnn}. In this work we
	deliberately pair two GNNs with different inductive biases, the
	pairwise MatErials Graph Network (MEGNet) and the three-body
	Atomistic Line Graph Neural Network (ALIGNN), which additionally
	encodes bond-angle triplets, so that requiring agreement between
	their predictions reduces the influence of any single model's
	systematic error~\cite{chen2019graph,choudhary2021atomistic}. As
	both are graph neural networks trained on the same data, they are
	not fully independent, and this consensus mitigates rather than
	eliminates model bias. We therefore add a third surrogate of a different
	kind, the universal machine-learning force field
	MatterSim~\cite{yang2024mattersim}, which is trained on independent
	PBE-level data rather than on adsorption energies and evaluates
	$\Delta G_{\mathrm{H}^*}$ from explicit total-energy differences. The
	three surrogates play complementary roles: the two GNNs screen every
	slab$+$H configuration at negligible cost, their agreement guards against
	architecture-specific errors, and MatterSim guards against errors shared
	through the common training data before DFT is committed.
	To our knowledge, no prior work has combined these components into a
	pipeline that
	generates novel bimetallic bulk structures conditioned on
	catalytic labels, constructs realistic slab$+$H libraries, and
	screens them with independently validated GNN surrogates. This
	work fills that gap. Our specific contributions are:
	(i) a MatterGen fine-tuning protocol on 1{,}569 bulk
	alloy HER labels whose 3{,}000 generated samples, after
	stable-and-unique filtering and restriction to database-novel
	binary compositions built from the training-set metals, yield
	730 novel bimetallic candidates;
	(ii) a MatterSim surface-energy screen that ranks the (100),
	(110), and (111) facet terminations of each bulk and retains the
	three most stable, from which a 39{,}132-configuration slab$+$H
	library is built by enumerating on-top, bridge, and hollow
	adsorption sites at a coverage of up to $0.25$~monolayer (ML) and relaxing each with the MatterSim
	machine-learning force field~\cite{yang2024mattersim};
	(iii) two independently trained GNN surrogates, MEGNet
	and ALIGNN, trained on the Mamun dataset to comparable
	accuracy (MAE 0.174 and 0.161~eV, $R^2=0.850$ and 0.879) on an
	identical held-out test set; (iv) a dual-model consensus screening on the free-energy scale
	(inter-model Pearson correlation coefficient $r=0.847$) that identifies 77
	near-thermoneutral bulk alloys
	($|\Delta G_{\mathrm{H}^*}|\le 0.2$~eV, 10.5\% of the 730 novel
	alloys), 50 with sign agreement and 49 after excluding
	non-converged and unphysical surfaces; and (v) plane-wave DFT
	validation of eight screened candidates,
	cross-checked by an orthogonal MatterSim energy-difference
	triangulation.
	
	The remainder of the paper is organised as follows.
	Section~\ref{sec:theory} reviews the HER free-energy
	descriptor that serves as the activity criterion.
	Section~\ref{sec:methods} details the five-stage computational
	pipeline. Section~\ref{sec:results} presents the generation and selection
	of the novel alloys, surrogate benchmarking, high-throughput screening
	and consensus analysis, the MatterSim triangulation, and the DFT and
	phonon validation of the selected candidates. Section~\ref{sec:conc} summarises
	conclusions and outlines directions for experimental
	follow-up.
	
	\section{HER Free-Energy Descriptor}
	\label{sec:theory}
	
	The electrochemical \HER{} proceeds through the
	Volmer--Heyrovsky or Volmer--Tafel mechanism, both sharing
	the hydrogen adsorption elementary step
	\begin{equation}
		\mathrm{H}^{+} + e^{-} + {*}
		\;\rightleftharpoons\;
		\mathrm{H}^{*},
		\label{eq:her}
	\end{equation}
	where $*$ denotes a surface site.
	The adsorption free energy
	$\Delta G_{\mathrm{H}^*} = \Delta E_{\mathrm{H}^*}
	+ \Delta E_{\mathrm{ZPE}} - T\Delta S$,
	where $\Delta E_{\mathrm{H}^*}$ is the electronic hydrogen adsorption
	energy, $\Delta E_{\mathrm{ZPE}}$ the change in zero-point energy,
	$T$ the temperature and $\Delta S$ the entropy change on adsorption,
	is the primary computational activity descriptor on the
	volcano plot~\cite{norskov2004origin,norskov2005trends}.
	An optimal catalyst satisfies
	$|\Delta G_{\mathrm{H}^*}| \approx 0$, ensuring neither
	adsorption nor desorption is rate-limiting.
	The thermoneutral window $|\Delta G_{\mathrm{H}^*}|
	\le 0.2$~eV is broadly accepted as the criterion for
	competitive activity at room
	temperature~\cite{greeley2006computational}.
	
	The zero-point and entropic contributions for adsorbed H are nearly
	constant across transition-metal surfaces, so that
	$\Delta G_{\mathrm{H}^*} = \Delta E_{\mathrm{H}^*} + 0.24$~eV within
	the computational hydrogen electrode (CHE)
	framework~\cite{norskov2005trends}. The GNN training labels in this
	work are electronic reaction energies $\Delta E_{\mathrm{H}^*}$ for
	$\tfrac12\mathrm{H_2(g)} + {*} \rightarrow \mathrm{H}^*$, computed
	with the Bayesian error estimation functional with van der Waals
	correlation (BEEF-vdW)~\cite{wellendorff2012density, perdew1996generalized} and reported
	without zero-point or entropy corrections~\cite{mamun2019high}. All GNN
	predictions are therefore converted to free energies with the same
	$+0.24$~eV term that is applied to the MatterSim and DFT energies
	(Eq.~\eqref{eq:CHE}).
	
	\section{Computational Methodology}
	\label{sec:methods}
	
\begin{figure*}[t]
		\centering
		\includegraphics[width=\textwidth]{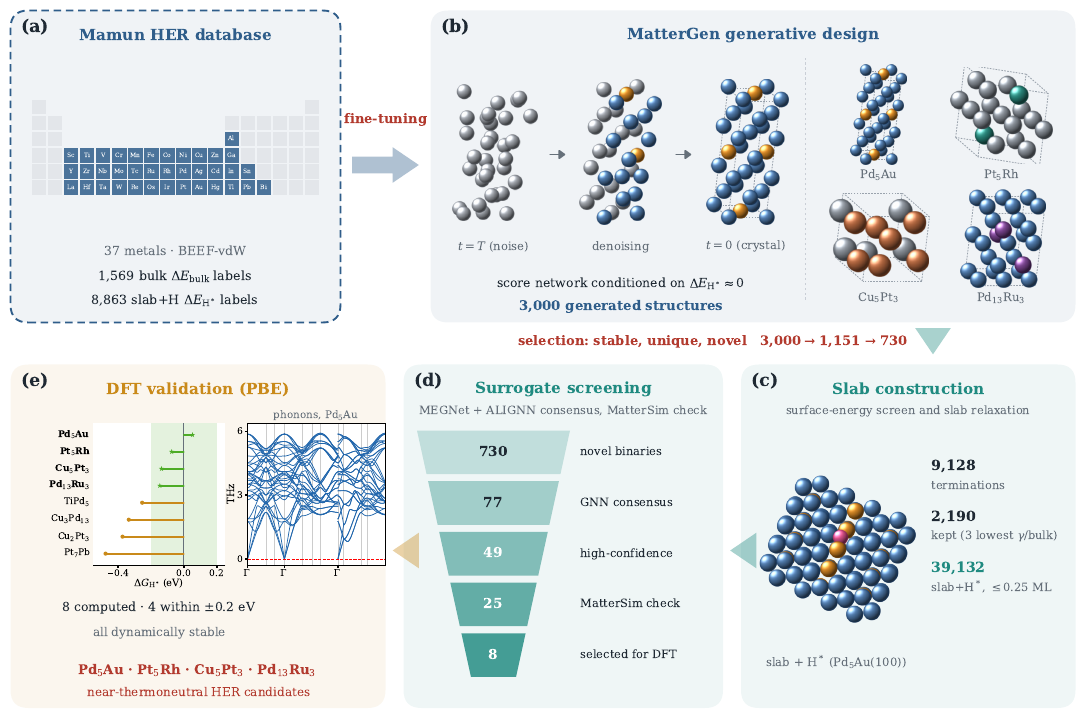}
		\caption{Discovery workflow. (a) MatterGen is fine-tuned on Mamun
			hydrogen adsorption energies for 37 metals. (b) The fine-tuned model
			generates 3{,}000 bulk structures by denoising, conditioned on
			$\Delta E_{\mathrm{H}^*}\approx0$, and 730 stable, unique, and
			database-novel binary alloys are selected. (c) MatterSim screens the
			surface terminations and relaxes 39{,}132 slab$+$H configurations.
			(d) The MEGNet and ALIGNN consensus, checked against MatterSim, narrows
			the candidates to eight. (e) DFT (PBE) and MatterSim phonons identify
			four near-thermoneutral, dynamically stable alloys.}
		\label{fig:workflow}
	\end{figure*}

	The workflow is illustrated schematically in
	Fig.~\ref{fig:workflow}.
	
	\subsection{Training Data and MatterGen Fine-Tuning}
	\label{sec:dataset}\label{sec:mattergen}
	
	All training data are taken from the Mamun
	high-throughput bimetallic dataset~\cite{mamun2019high},
	which provides DFT hydrogen chemisorption energies
	for 1{,}998 bimetallic and 37 pure-metal surfaces
	ranging over three Strukturbericht designations
	(A1, L1$_2$, L1$_0$) across 37 metals, each on its close-packed
	facet.
	All calculations use Quantum ESPRESSO~\cite{giannozzi2017quantum}
	with the BEEF-vdW functional~\cite{wellendorff2012density},
	a 500~eV plane-wave cutoff, $(6,6,1)$ Monkhorst--Pack
	$k$-point grids, and three-layer $(2\times2)$ slabs with the two
	bottom layers fixed.
	
	For generative fine-tuning we build a bulk dataset of 1{,}569 unique
	bulk alloy structures, one per distinct chemical composition
	in the Mamun dataset, each labelled with the minimum reaction
	energy across all available adsorption sites and facets:
	\begin{equation}
		\Delta E_{\mathrm{bulk}} =
		\min_{\alpha,\,f}\;\Delta E_{\mathrm{H}^*}^{(\alpha,f)},
		\label{eq:gmin}
	\end{equation}
	where $\alpha$ indexes adsorption site and $f$ the facet
	(a single close-packed facet per structure in Mamun).
	This scalar encodes the intrinsic catalytic potential of
	each bulk composition while remaining fully compatible
	with the MatterGen bulk-structure generation.
	The fine-tuning labels range from $-2.75$ to $+2.74$~eV
	(mean $-0.65$~eV).
	
	MatterGen~\cite{zeni2025generative} is a denoising
	diffusion model that jointly learns score functions for
	atom types, fractional coordinates, and the periodic
	lattice through an equivariant message-passing backbone.
	Property-conditional generation is implemented via
	adapter modules~\cite{zhang2023conditional} injected at
	each network layer and trained on labelled (structure,
	property) pairs, with generation steered by
	classifier-free guidance~\cite{ho2022classifier}.
	MatterGen is a bulk-crystal generator that enforces
	three-dimensional periodicity, so it is not suited to producing
	the slab geometries (with vacuum and a broken periodic direction)
	needed for adsorption screening. We therefore separate the two
	tasks: MatterGen generates bulk alloys, and the slabs are built
	from these bulks in a subsequent step (Section~\ref{sec:slab}).
	Accordingly, MatterGen is fine-tuned on the 1{,}569-structure bulk
	dataset and then conditioned on $\Delta E_{\mathrm{bulk}}\approx0$~eV to
	generate 3{,}000 bulk structures. This target corresponds to
	$\Delta G_{\mathrm{H}^*}\approx+0.24$~eV (Section~\ref{sec:theory}), and
	the screening that follows is carried out on $\Delta G_{\mathrm{H}^*}$.
	The selection of the novel bimetallic alloys from the generated
	structures is described in Section~\ref{sec:gen_res}.
	
	For the GNN surrogates we extract the slab$+$H subset of the Mamun
	dataset, retaining structures
	with $-5 \le \Delta E_{\mathrm{H}^*} \le 5$~eV and
	available crystallographic information file (CIF) strings, resulting in 8{,}863 entries
	ranging $-2.64$ to $+3.15$~eV
	(mean $-0.211$~eV, std~0.793~eV).
	The dataset is partitioned into training (6{,}203),
	validation (1{,}330), and test (1{,}330) sets with a
	fixed random seed (42).
	
	\subsection{Slab and Adsorption-Site Construction}
	\label{sec:slab}
	
	Surface slabs were generated from each of the 730 novel bimetallic
	bulk structures (Section~\ref{sec:gen_res}), first converted to their conventional standard cells
	so that the Miller indices identify well-defined crystallographic
	surfaces. For each bulk the (100), (110), and (111) orientations were
	cut with a minimum slab thickness of 8~\AA{} and a vacuum spacing of
	15~\AA{}, and all symmetrically distinct terminations along
	each orientation were retained rather than a single arbitrary cut.
	
	Because a low-symmetry intermetallic can expose many terminations,
	many of which cut through strong bonds and are energetically unfavourable,
	we rank the terminations by an approximate surface energy. Each clean
	(adsorbate-free) slab is relaxed with MatterSim and scored by
	$\gamma \approx [E_\mathrm{slab} - N\,E_\mathrm{bulk}]/2A$, where
	$E_\mathrm{slab}$ is the total energy of the relaxed slab, $N$ its number
	of atoms, $E_\mathrm{bulk}$ the bulk energy per atom, and $A$ the area of
	one slab face (the factor 2 counts the two faces). The relation is
	approximate because the two faces of a slab are generally inequivalent,
	so $\gamma$ is their mean.
	Only
	the three lowest-$\gamma$ (facet, termination) pairs per bulk are
	carried forward. This step both discards physically irrelevant
	high-energy surfaces and normalises the number of terminations per
	composition, preventing compositions that happen to enumerate many
	cuts from dominating the subsequent
	minimum-$\Delta G_{\mathrm{H}^*}$ screen. The screen relaxed 9{,}128
	clean slabs and retained 2{,}190 surfaces (three per bulk), whose
	surface energies range $0.09$--$2.54$~J\,m$^{-2}$.
	
	On each retained surface the in-plane cell was expanded so that the
	top atomic layer contains at least four atoms, giving a hydrogen
	coverage of at most $0.25$~ML, the coverage of the Mamun $(2\times2)$
	training slabs~\cite{mamun2019high}. In all, 76\% of the configurations are at
	exactly $0.25$~ML and the remainder, on surfaces with larger top-layer
	cells, at $0.08$--$0.20$~ML. Hydrogen adsorption sites were then
	enumerated at all symmetry-inequivalent high-symmetry positions:
	on-top (above a surface atom), bridge (midpoint between two
	neighbouring surface atoms), and hollow (centre of a three- or
	four-fold ring of surface atoms). Symmetry-equivalent and
	near-duplicate sites were merged, geometrically unphysical obtuse
	hollow sites were discarded, and a single hydrogen atom was placed
	1.5~\AA{} above each retained site. Sampling on-top, bridge, and
	hollow sites matches the site enumeration of the Mamun training
	dataset~\cite{mamun2019high}, keeping the screening input in the same
	distribution as the GNN training data.
	
	This procedure yielded 39{,}132 slab$+$H configurations across the
	730 bulk structures. Each was relaxed with the MatterSim universal
	machine-learning force field (MLFF)~\cite{yang2024mattersim} using the Broyden--Fletcher--Goldfarb--Shanno (BFGS)
	optimiser to a maximum residual atomic force of $f_\mathrm{max}=0.05$~eV/\AA{}
	(maximum 500 ionic steps). By analogy with the Mamun slabs, in which
	only the top layer and the adsorbate relax, only the
	adsorbed hydrogen and the metal atoms within $2$~\AA{} of the surface
	were relaxed while the deeper bulk-like atoms were held fixed at
	their standardised positions. The 255 relaxations (0.7\%) that did not
	reach the force criterion within 500 steps were kept in the screen and
	flagged as non-converged (Section~\ref{sec:consensus}).
	
	\subsection{GNN Surrogate Models}
	\label{sec:gnn}
	
	We screen with two graph neural network surrogates that encode the
	local adsorption environment through complementary channels. A
	structure is represented as an attributed graph whose nodes carry
	elemental features and whose edges carry interatomic bonds, and
	each model predicts $\Delta G_{\mathrm{H}^*}$ by iterative message
	passing, without hand-crafted descriptors.
	MEGNet~\cite{chen2019graph} is a two-body model whose edges carry
	Gaussian-expanded interatomic distances, so bond angles enter only
	implicitly. ALIGNN~\cite{choudhary2021atomistic} adds explicit
	three-body information by also passing messages on the line graph,
	whose edges represent bond-angle triplets. Requiring the two
	models to agree therefore reduces the influence of either
	inductive bias (Section~\ref{sec:bench}).
	
	MEGNet (189{,}521 trainable parameters) is initialised from a
	checkpoint pretrained on Materials Project formation
	energies~\cite{jain2013materials,chen2019graph}. We fine-tune all
	pretrained weights, including those of the output layers, for scalar
	adsorption-energy regression, and
	replace the zero-initialised global state vector with a
	two-dimensional composition descriptor selected by Pearson
	correlation against the training labels: the mean number of
	unfilled $d$-orbitals per atom ($|\rho|=0.658$) and the mean
	Mendeleev number ($|\rho|=0.616$), both $z$-score normalised
	using training-set statistics. Fine-tuning uses the AdamW
	optimiser~\cite{loshchilov2017decoupled} (learning rate
	$5\times10^{-4}$, weight decay $10^{-4}$, batch size 32), a Huber
	loss ($\delta=0.5$), a cosine-annealing schedule, dropout 0.15 on
	the output multilayer perceptron, and early stopping on validation MAE (patience 25,
	up to 200 epochs). Its bond-graph cutoff is extended to 5.0~\AA{}
	(from 4.0~\AA{}).
	
	ALIGNN (4{,}026{,}753 trainable parameters) is trained from random
	initialisation with four ALIGNN
	layers and four graph convolution layers, hidden dimension 256,
	embedding dimension 64, and 92-dimensional crystal graph convolutional neural network
	(CGCNN) atom features~\cite{xie2018crystal}. Its bond-graph cutoff of 8.0~\AA{}
	follows the ALIGNN default for crystals, broad enough that
	under-coordinated surface atoms still form sufficient line-graph
	edges for stable triplet message passing. Training uses the AdamW
	optimiser~\cite{loshchilov2017decoupled} (learning rate
	$10^{-3}$, weight decay $10^{-4}$, batch size 32), a mean-squared-error loss,
	cosine annealing, gradient clipping (max norm 5.0), and early
	stopping on validation MAE (patience 30, up to 300 epochs).
	
	\subsection{Screening Protocol}
	\label{sec:screen_prot}
	
	The trained ALIGNN and MEGNet models each predict
	$\Delta E_{\mathrm{H}^*}$ for all 39{,}132 MatterGen slab$+$H
	configurations, converted to
	$\Delta G_{\mathrm{H}^*} = \Delta E_{\mathrm{H}^*} + 0.24$~eV
	(Section~\ref{sec:theory}). For each bulk material and each model, the adsorption site
	with the smallest $|\Delta G_{\mathrm{H}^*}|$ across all facets
	and sites is retained as the representative descriptor, so the two
	models can select different sites.
	Candidates are ranked at three thresholds:
	$|\Delta G_{\mathrm{H}^*}| \le 0.1$~eV (excellent),
	$\le 0.2$~eV (good), and $\le 0.3$~eV (acceptable).
	Consensus candidates are then defined as materials
	whose larger single-model prediction satisfies
	$\max(|\Delta G_\mathrm{ALIGNN}|,|\Delta G_\mathrm{MEGNet}|)
	\le\varepsilon$; sign agreement between the two models is
	tracked as an additional confidence criterion.
	
	\subsection{MatterSim Energy-Difference Pre-Validation}
	\label{sec:mlff_prevalidation}
	
	Before committing plane-wave DFT resources, the shortlisted
	consensus candidates were pre-validated using MatterSim total
	energies within the CHE
	framework~\cite{norskov2004origin}:
	\begin{equation}
		\Delta G_{\mathrm{H}^*} = E(\text{slab}{+}\text{H}) - E(\text{slab})
		- \tfrac{1}{2}E(\mathrm{H_2}) + 0.24~\text{eV},
		\label{eq:CHE}
	\end{equation}
	where $E(\text{slab}{+}\text{H})$, $E(\text{slab})$ and
	$E(\mathrm{H_2})$ are the total energies of the slab with adsorbed H,
	the clean slab and a gas-phase H$_2$ molecule, and the 0.24~eV term captures the zero-point and entropic
	corrections for adsorbed H at 298~K~\cite{norskov2005trends}.
	This step is method-orthogonal to the GNN screening: whereas
	the GNNs regress $\Delta G_{\mathrm{H}^*}$ directly from
	structure, Eq.~\eqref{eq:CHE} reconstructs it from independent
	MLFF total-energy differences. For each candidate the ALIGNN best-site configuration was
	evaluated. ALIGNN was used because it has the lower test-set error of
	the two GNNs, and a single model's site has to be chosen
	because the two GNNs place their best sites on the same configuration
	for only 74 of the 730 bulks (Section~\ref{sec:consensus}). Bare-slab energies were obtained by removing the adsorbed
	H and re-relaxing to $f_\mathrm{max}=0.05$~eV/\AA{}, and the H$_2$
	reference ($E(\mathrm{H_2})=-6.735$~eV) was computed for an isolated
	molecule in a $12\times12\times12$~\AA$^3$ box.
	
	\subsection{First-Principles DFT Validation}
	\label{sec:dft_meth}
	
	Eight candidates of distinct chemistry
	(Pd$_5$Au, Pd$_{13}$Ru$_3$, TiPd$_5$, Cu$_3$Pd$_{13}$, Pt$_5$Rh,
	Cu$_5$Pt$_3$, Cu$_2$Pt$_3$, and Pt$_7$Pb), selected as described in
	Section~\ref{sec:dft_res}, were validated with plane-wave
	DFT using Quantum ESPRESSO~\cite{giannozzi2017quantum}.
	Calculations employed the PBE exchange--correlation
	functional~\cite{perdew1996generalized} with projector augmented-wave (PAW) pseudopotentials
	from the \texttt{pslibrary}, plane-wave kinetic-energy cutoffs of
	$E_\mathrm{cut}=60$--$90$~Ry and charge-density cutoffs of
	$520$--$720$~Ry. Metallic occupations were treated with cold
	smearing~\cite{marzari1999thermal} ($\mathrm{degauss}=0.02$~Ry), and
	the Brillouin zone was sampled with $\Gamma$-centred Monkhorst--Pack
	grids of $(4$--$11)\times(4$--$11)\times1$, with only the in-plane
	directions sampled owing to the $>15$~\AA{} vacuum. Consistent with
	the screening protocol, only the adsorbed hydrogen and the metal
	atoms within $2$~\AA{} of the surface were relaxed while the deeper
	atoms were held fixed (BFGS, force threshold $10^{-4}$~Ry/bohr),
	ensuring a consistent reference between the bare slab and the
	slab$+$H system. Spin-unpolarised calculations were used, as none
	of the eight candidates contains a magnetic element. The gas-phase H$_2$ reference ($E(\mathrm{H_2})=-31.75$~eV) was
	computed at $90$/$720$~Ry in a $15\times15\times15$~\AA$^3$ box with
	fixed occupations and $\Gamma$-point sampling. Free energies were obtained from
	Eq.~\eqref{eq:CHE}.
	
	\subsection{Phonon Calculations}
	\label{sec:phonon_meth}
	
	The dynamical stability of the DFT candidates was assessed from phonon
	dispersions computed by the finite-displacement method in
	Phonopy~\cite{togo2015first}, with interatomic forces from the same
	MatterSim force field~\cite{yang2024mattersim} used throughout the
	pipeline. Each bulk cell was relaxed (cell and internal coordinates,
	$f_\mathrm{max}=0.01$~eV/\AA{}) and displaced by $0.03$~\AA{} in
	$64$--$256$-atom supercells, and the resulting force constants were used
	to compute the dispersion along the standardised high-symmetry
	path~\cite{hinuma2017band} and the element-projected phonon density of
	states.
	
	\section{Results and Discussion}
	\label{sec:results}
	
	\subsection{Generation and Selection of Novel Bimetallic Alloys}
	\label{sec:gen_res}
	
	Each of the 3{,}000 generated structures was pre-relaxed with MatterSim
	and classified as stable if its relaxed energy per atom lies within
	0.1~eV/atom of the reference convex hull, novel if it is absent from the
	Alexandria and Materials Project (Alex-MP) reference set released with
	MatterGen, and unique if no other generated structure is
	symmetry-equivalent to it.  Of the 3{,}000 structures,
	1{,}374 (45.8\%) are stable, 2{,}197 (73.2\%) novel, and 2{,}651
	(88.4\%) unique. In total, 1{,}151 (38.4\%) are both
	stable and unique, and 673 (22.4\%) are stable, unique, and novel (SUN).
	The high novelty and uniqueness rates show that the model explores new
	chemical compositions rather than reproducing known ones.
	
	\begin{table}[h]
		\centering
		\caption{Composition types of the 3{,}000 MatterGen-generated
			structures. ``In-distribution'' means all elements are present in the
			1{,}569-sample Mamun fine-tuning set.}
		\label{tab:sun}
		\small
		\setlength{\tabcolsep}{4.5pt}
		\begin{tabular}{lrrl}
			\toprule
			Type & Count & \% & In-distribution \\
			\midrule
			Bimetallic    & 2330 & 77.7 & 2298 (98.6\%) \\
			Ternary       &  465 & 15.5 & 379 (81.5\%) \\
			4-component   &   88 &  2.9 & 38 (43.2\%) \\
			Monometallic  &   66 &  2.2 & 66 (100\%)  \\
			5-component   &   29 &  1.0 & 6 (20.7\%)  \\
			6+ components &   22 &  0.7 & 0 (0\%)     \\
			\midrule
			Total & 3000 & 100 & 2725 (90.8\%) \\
			\bottomrule
		\end{tabular}
	\end{table}
	
	Table~\ref{tab:sun} breaks the generated structures down by
	composition. Most are bimetallic (77.7\%), consistent with the binary
	training set, but 15.5\% are ternary, 2.9\% four-component, and smaller
	fractions extend to nine-component compositions, while 2.2\% are
	monometallic. MatterGen's base model was pretrained on the Alex-MP-20
	database of multicomponent inorganic solids, and fine-tuning on 1{,}569
	binary alloy labels steers the generative distribution without imposing
	a hard two-element constraint during sampling. The model therefore also
	produces stoichiometries absent from the fine-tuning set, such as
	ZnNi$_6$Mo, Cu$_{11}$Pt$_6$Au$_3$, and Zn$_2$Co$_3$Cu$_5$. Overall,
	2{,}725 structures (90.8\%) contain only elements present in the Mamun
	fine-tuning set. Of the remaining 275, 232 contain non-metals or metalloids
	(most often H, in 103 structures, followed by I, Te, S, C, Se, N, and B),
	typical of the chalcogenide, halide, and carbide phases of the
	pretraining corpus, and 75 contain alkali, alkaline-earth, or lanthanide
	metals absent from the training set (most often Cs, Be, Rb, and K), with
	some structures in both groups.
	
	Both GNN surrogates were trained exclusively on pure-metal and
	bimetallic slab$+$H data, so their predictions for higher-component
	structures, or for elements outside the training set, would be
	out-of-distribution extrapolations. The screen is therefore restricted
	to binary alloys within the training domain. From the 1{,}151 stable and
	unique structures we retain only binary alloys whose two elements both
	belong to the Mamun metal set, whose stoichiometry differs from the pure,
	A$_3$B, and AB ratios that Mamun tabulates, and whose relaxed structure
	matches no Mamun prototype under a StructureMatcher comparison. This
	yields the 730 novel bimetallic alloys (24.3\% of the generated
	structures, 680 distinct formulas) screened in this work, all absent
	from the training set but within its elemental domain. Their novelty is
	assessed against the Mamun training set, the relevant reference for the
	surrogates, so this set differs from the 673 SUN structures, whose
	novelty is assessed against Alex-MP. Of the 730, 500 are also
	novel with respect to Alex-MP, and the other 230 match an entry of
	that reference set but have no hydrogen-adsorption data in Mamun. For
	these 230 the new information is the predicted HER activity rather than
	the structure. Throughout this work, novel refers to novelty with respect
	to the Mamun dataset. Judged by the space group of the metal sublattice with
	chemical identity ignored, 279 of the 730 (38\%) are chemically ordered
	superstructures on an fcc parent lattice and 451 (62\%) are built on
	other lattices. Indeed, as shown in Fig.~\ref{fig:S1}, among the four candidates validated by DFT, Pd$_{13}$Ru$_{3}$ ($P2$, no. 3), Pd$_{5}$Au ($Cmmm$, no. 65), and Pt$_{5}$Rh ($P\bar{3}m1$, no. 164) follow an ABC stacking of fcc, whereas Cu$_{5}$Pt$_{3}$ ($Pm$, no.~6) follows an AB stacking of hcp. Except for Cu$_{5}$Pt$_{3}$, this is consistent with the experimental binary phase diagrams, which report fcc solid solutions at these compositions \cite{asm2016}, whereas no hcp phase is known in the Cu--Pt system. This discrepancy reflects a known limitation of MatterGen and, more generally, of current generative models for crystals \cite{zeni2025generative, Leeman2024} that cannot represent partially occupied site. As a consequence, an ordered approximant of a known disordered phase can be classified as novel unless the comparison with reference databases accounts for compositional disorder \cite{Juelsholt2026}. Experimentally, the compositions considered here are expected to form chemically disordered solid solutions rather than ordered intermetallics. In the remainder of this article, the ordered generated structures are nevertheless retained as structural models, since DFT requires an ordered cell. Each of them should be regarded as one particular local chemical environment of the corresponding solid solution.\\
	
\begin{figure*}[t]
		\centering
		\includegraphics[width=0.92\textwidth]{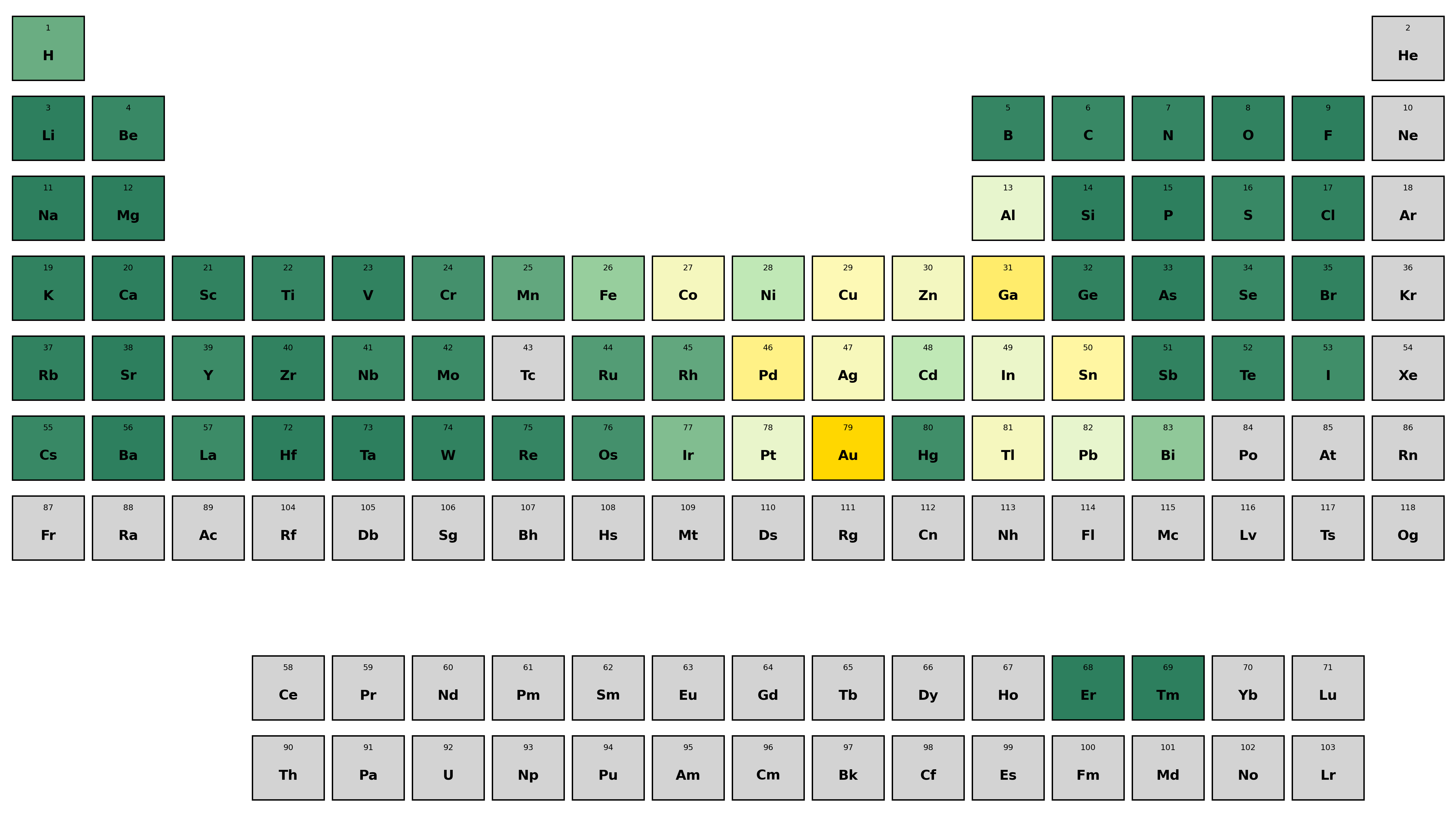}
		\caption{Occurrence frequency of each element across the 3{,}000
			MatterGen-generated bulk structures (yellow: highest frequency; dark
			green: low but present; light grey: absent).}
		\label{fig:heatmap}
	\end{figure*}

	
	The elemental distribution across all 3{,}000 generated structures is
	shown in Fig.~\ref{fig:heatmap}. Au is the single
	most-generated element (606 structures), and Pd--Au, Ag--Au, Au--Cu, and
	Au--Zn are among the top generated pairs. Ga (455 structures), Pd (428),
	and Sn (398) follow, with the post-transition metals In, Tl, and Pb also
	prominent. The earth-abundant late $d$-block metals Co, Ni, Cu, and Zn
	appear at moderate frequency (242--374 structures each), while the early
	$d$-block metals Ti, V, Cr, and Mn are generated rarely (11--87
	structures each). This pattern
	follows the fine-tuning labels. Bulks containing Au, Ga, In, Sn, or Pb
	have the labels closest to zero (median $\Delta E_{\mathrm{bulk}}$
	between $-0.04$ and $+0.08$~eV), whereas those containing Ti, V, or Cr
	bind hydrogen strongly (medians $-0.9$ to $-1.1$~eV), on the strongly
	binding flank of the HER volcano~\cite{norskov2005trends}. The
	suppression of strongly binding elements is therefore consistent with
	the conditioning signal reaching the generative distribution.
	
	\subsection{Surrogate Model Benchmarking}
	\label{sec:bench}
\begin{figure*}[t]
		\centering
		\includegraphics[width=0.75\textwidth]{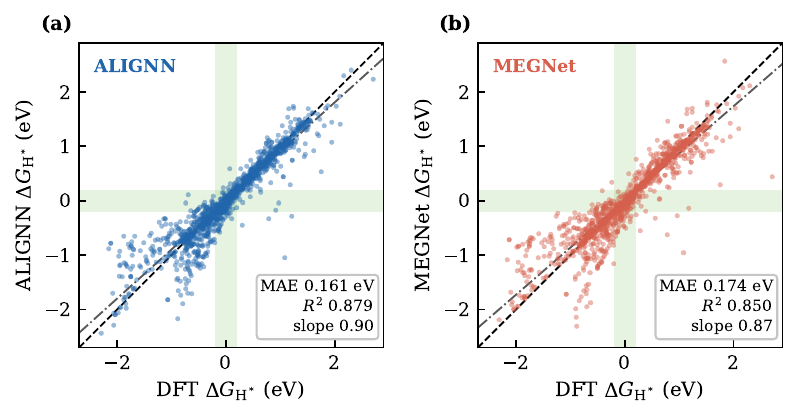}
		\caption{Surrogate benchmarking on the held-out Mamun test set
			($n_{\mathrm{test}}=1{,}330$). (a) ALIGNN and (b) MEGNet predictions
			against the DFT (BEEF-vdW) values, both on the free-energy scale
			$\Delta G_{\mathrm{H}^*}=\Delta E_{\mathrm{H}^*}+0.24$~eV, with the
			identity line (dashed) and a linear fit (dash-dotted). Green bands mark
			the thermoneutral window $|\Delta G_{\mathrm{H}^*}|\le0.2$~eV.}
		\label{fig:gnn}
	\end{figure*}

	Test-set metrics for both GNN models are summarised in
	Table~\ref{tab:metrics}, and the test-set parity plots are shown in
	Fig.~\ref{fig:gnn}.
	\begin{table}[h]
		\centering
		\caption{Test-set performance on the Mamun slab$+$H
			dataset ($n_{\mathrm{test}}=1{,}330$). RMSE: root-mean-square error.}
		\label{tab:metrics}
		\small
		\begin{tabular}{lccc}
			\toprule
			Model  & MAE (eV) & RMSE (eV) & $R^2$ \\
			\midrule
			MEGNet          & 0.174 & 0.302 & 0.850 \\
			ALIGNN          & 0.161 & 0.270 & 0.879 \\
			\bottomrule
		\end{tabular}
	\end{table}
	The two models reach closely comparable accuracy (MAE 0.174 and
	0.161~eV; $R^2$ 0.850 and 0.879). They encode the local adsorption environment
	differently, MEGNet through pairwise distances and ALIGNN through
	explicit bond-angle triplets, and we exploit this complementarity
	in the consensus analysis (Section~\ref{sec:consensus}) rather
	than selecting a single best model. The regression slopes, 0.87
	(MEGNet) and 0.90 (ALIGNN), indicate that both regress mildly
	toward the mean at the extremes of the energy range, as expected
	for finite training data.
	
	Because the screen selects candidates by whether
	$\Delta G_{\mathrm{H}^*}$ falls inside the thermoneutral window, we also
	evaluate the surrogates as classifiers for that window
	(Table~\ref{tab:window}). Of the 1{,}330 test structures, 298 (22.4\%)
	have DFT values within $\pm0.2$~eV. ALIGNN and MEGNet place 75.2\% and
	78.9\% of them inside the window (recall), and 80.0\% and 80.2\% of
	their in-window predictions are confirmed by DFT (precision). Within the
	window the MAE is about half its overall value (0.083 and 0.090~eV).
	Requiring both
	models to place a structure inside the window, the consensus criterion
	of the screen (Section~\ref{sec:screen_prot}), raises the precision to
	85.4\% at a recall of 66.8\%. For a screen that passes candidates to
	DFT, precision matters more than recall: missing some active structures
	costs less than computing false positives. Requiring in addition that
	the two models agree in sign leaves the precision almost unchanged
	(84.1\%) and lowers the recall to 51.3\%, but raises the fraction of
	selected structures whose predicted sign matches DFT from 81\% (ALIGNN)
	and 77\% (MEGNet) to 87\%. Sign agreement therefore indicates on which
	side of thermoneutral a candidate lies, not whether it lies inside the
	window.
	\begin{table}[h]
		\centering
		\caption{Thermoneutral-window classification on the held-out Mamun test
			set ($n_{\mathrm{test}}=1{,}330$, of which 298 have DFT (BEEF-vdW)
			$|\Delta G_{\mathrm{H}^*}|\le0.2$~eV). $N_\mathrm{pred}$: test
			structures predicted inside the window. Correct: predicted inside and
			DFT inside. Recall: correct/298. Precision: correct/$N_\mathrm{pred}$.
			MAE$_\mathrm{win}$: MAE over the 298 structures with DFT values inside
			the window. Consensus: both models predict inside the window.
			$+$ sign: consensus, and both predictions have the same sign.}
		\label{tab:window}
		\small
		\setlength{\tabcolsep}{3.5pt}
		\begin{tabular}{lrrrrr}
			\toprule
			Model & $N_\mathrm{pred}$ & Correct & Recall & Precision & MAE$_\mathrm{win}$ \\
			&                   &         & (\%)   & (\%)      & (eV) \\
			\midrule
			ALIGNN    & 280 & 224 & 75.2 & 80.0 & 0.083 \\
			MEGNet    & 293 & 235 & 78.9 & 80.2 & 0.090 \\
			Consensus & 233 & 199 & 66.8 & 85.4 & --    \\
			\quad $+$ sign & 182 & 153 & 51.3 & 84.1 & -- \\
			\bottomrule
		\end{tabular}
	\end{table}
	
	\subsection{High-Throughput Screening of MatterGen Candidates}
	\label{sec:screen_res}
	
\begin{figure}[h]
		\centering
		\includegraphics[width=\linewidth]{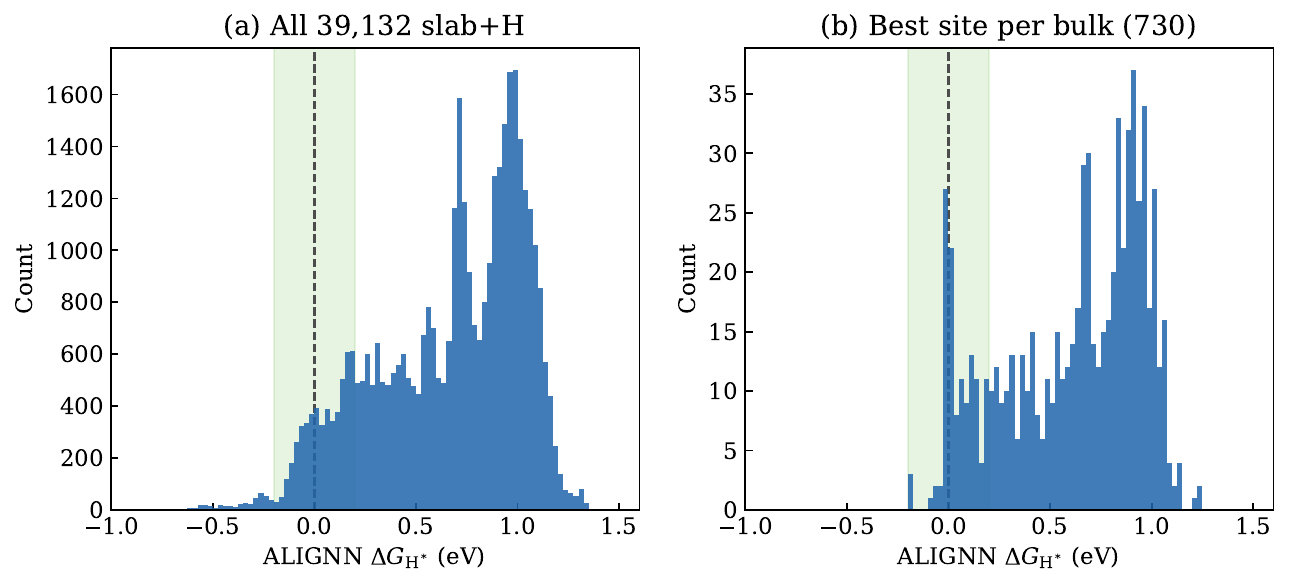}
		\caption{Distribution of ALIGNN-predicted
			$\Delta G_{\mathrm{H}^*}$.
			(a) All 39{,}132 slab$+$H configurations.
			(b) Best adsorption site per bulk material (730 total).
			Restricting to the optimal site enriches the
			thermoneutral window, raising the $|\Delta G|\le0.2$~eV
			fraction to 17.0\% (124 materials).
			Green shading: thermoneutral window $|\Delta G|\le0.2$~eV.
			Dashed vertical line: $\Delta G=0$.}
		\label{fig:screening}
	\end{figure}

	Both ALIGNN and MEGNet return predictions for all 39{,}132 slab$+$H
	configurations. All values below are free energies,
	$\Delta G_{\mathrm{H}^*}=\Delta E_{\mathrm{H}^*}+0.24$~eV
	(Section~\ref{sec:theory}). The distribution of ALIGNN-predicted
	$\Delta G_{\mathrm{H}^*}$ across all 39{,}132 slab$+$H
	configurations is shown in Fig.~\ref{fig:screening}(a).
	Of all 39{,}132 slab$+$H predictions, 2{,}743 fall within
	$|\Delta G_{\mathrm{H}^*}|\le 0.1$~eV (7.0\%), and 5{,}238
	within $\le 0.2$~eV (13.4\%).
	
	Retaining only the best-site prediction per bulk material
	(i.e.\ the adsorption site with the smallest
	$|\Delta G_{\mathrm{H}^*}|$ across all facets and sites)
	enriches the thermoneutral fraction, as shown in
	Fig.~\ref{fig:screening}(b) and quantified in
	Table~\ref{tab:hits}.
	For ALIGNN, the best-site $\le0.2$~eV fraction is 17.0\% (124 of
	730 bulks), and MEGNet gives a comparable best-site rate of 19.5\%
	(142 of 730).
	
	\begin{table}[h]
		\centering
		\caption{GNN best-site screening hit rates: number (and
			fraction) of bulk materials whose best adsorption site lies
			within each $|\Delta G_{\mathrm{H}^*}|$ threshold.
			ALIGNN and MEGNet each score all 730 novel bimetallic bulks.
			The consensus column requires
			$\max(|\Delta G_\mathrm{ALIGNN}|,|\Delta G_\mathrm{MEGNet}|)$
			within threshold.}
		\label{tab:hits}
		\small
		\setlength{\tabcolsep}{3.1pt}
		\begin{tabular}{lrrrrrr}
			\toprule
			Thresh. & \multicolumn{2}{c}{ALIGNN (730)}
			& \multicolumn{2}{c}{MEGNet (730)}
			& \multicolumn{2}{c}{Consensus (730)} \\
			\cmidrule(lr){2-3}\cmidrule(lr){4-5}\cmidrule(lr){6-7}
			(eV) & $N$ & \% & $N$ & \% & $N$ & \% \\
			\midrule
			$\le 0.1$ &  82 & 11.2 &  93 & 12.7 &  43 &  5.9 \\
			$\le 0.2$ & 124 & 17.0 & 142 & 19.5 &  77 & 10.5 \\
			$\le 0.3$ & 165 & 22.6 & 186 & 25.5 & 123 & 16.8 \\
			\bottomrule
		\end{tabular}
	\end{table}
	
	\subsection{Dual-Model Consensus and Prediction Reliability}
	\label{sec:consensus}
	
	\subsubsection{Inter-model agreement}
	
	Despite their different inductive biases and independent
	training (Section~\ref{sec:gnn}), the best-site predictions
	of the two models are well correlated across all 730
	novel bimetallic alloys
	(Pearson $r = 0.847$,
	Fig.~\ref{fig:consensus}(a)), and their predictions for individual
	configurations correlate almost as well ($r=0.790$ over all 39{,}132).
	The two models therefore largely agree, and the residual scatter
	identifies the alloys on which they disagree. Each model's best site is
	selected independently, and the two fall on the same configuration for
	only 74 of the 730 bulks, because a near-thermoneutral bulk typically
	offers many sites predicted close to zero. The consensus is therefore a
	criterion on the material, requiring both models to find a
	near-thermoneutral site on it, rather than on a single site.
	
	\subsubsection{Consensus candidates}
	
\begin{figure}[h]
		\centering
		\includegraphics[width=\linewidth]{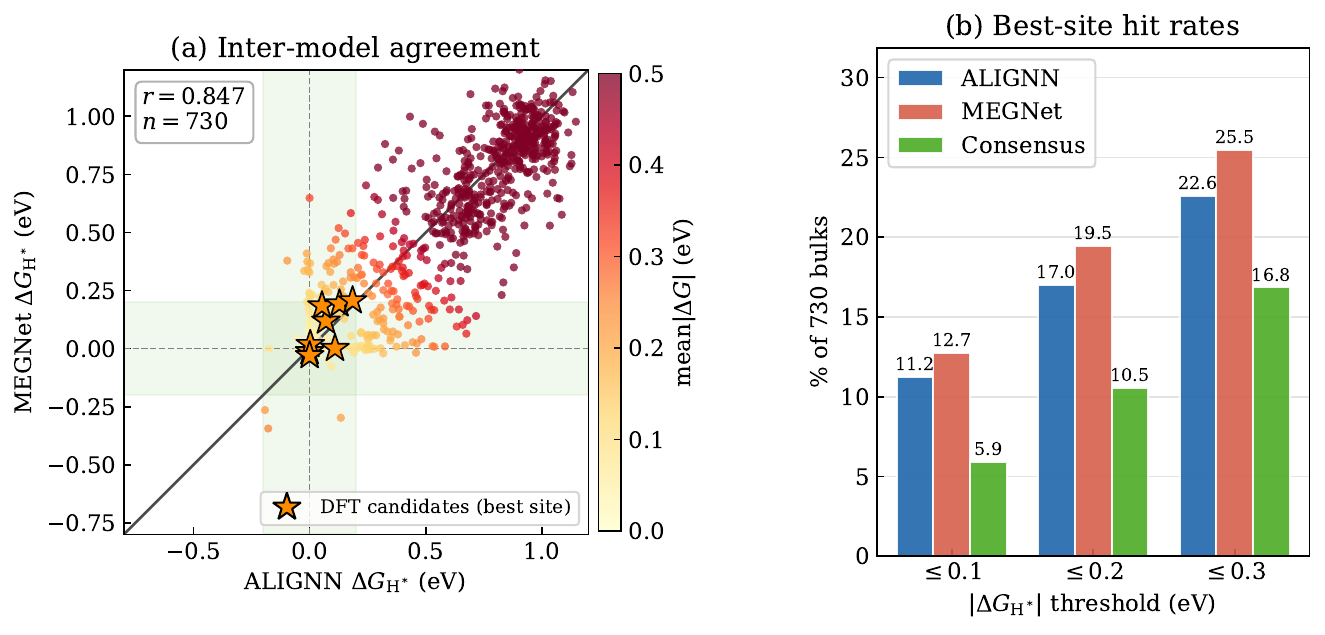}
		\caption{(a) ALIGNN vs.\ MEGNet best-site
			$\Delta G_{\mathrm{H}^*}$ for all 730 novel bimetallic bulk
			materials, coloured by mean $|\Delta G|$. Green bands
			mark the $\pm0.2$~eV thermoneutral window, and the eight
			DFT candidates are starred at their best sites. The inter-model
			Pearson correlation is $r=0.847$. (b) Best-site hit rates at three
			thresholds for ALIGNN, MEGNet, and their consensus, over the
			730 novel alloys.}
		\label{fig:consensus}
	\end{figure}

	The consensus set, defined by
	$\max(|\Delta G_\mathrm{ALIGNN}|,|\Delta G_\mathrm{MEGNet}|)
	\le 0.2$~eV, contains 77 bulk materials
	(10.5\% of the 730 novel alloys). For 71 of them at least one
	configuration lies inside the window in both models, and the consensus
	materials have a median of 40 (ALIGNN) and 11 (MEGNet) in-window
	configurations each. At the stricter
	$\le0.1$~eV threshold, 43 materials (5.9\%) qualify
	(Table~\ref{tab:hits}, Fig.~\ref{fig:consensus}(b)).
	Of the 77 consensus candidates, 50 additionally
	exhibit sign agreement between the two models, and 49 remain
	after excluding Cd$_2$Pd$_{13}$, whose best-site relaxation did not
	converge. These 49 form the highest-confidence subset for validation. A
	cut on unphysically low surface energies
	($\gamma<0.3$~J\,m$^{-2}$) removes none of them. The consensus set is enriched in fcc-derived
	superstructures: 54 of the 77 (70\%) and 34 of the 49
	high-confidence members, against 38\% of all 730 novel alloys.
	
	Of the single-model near-thermoneutral candidates, 62\% (ALIGNN) and
	54\% (MEGNet) are also placed in the window by the other model. The two
	models select the same configuration for only 2 of the 77 consensus
	alloys, but 71 of them have at least one configuration that both place
	within the window. Agreement between the two GNNs cannot remove a bias
	they share through the common training data, which the MatterSim
	comparison addresses (Section~\ref{sec:mlff_res}). The 77 consensus
	alloys, and first the 49 high-confidence ones, are passed to the
	validation steps below.
	
	\subsubsection{Top consensus candidates}
	
	The top-ranked consensus candidates (Table~\ref{tab:top20}) are mostly
	alloys of the Pt-group metals (Pt, Ir, Pd) with Au, Ag, Cu, In, and Pb,
	consistent with the near-thermoneutral hydrogen binding of Pt-group
	surfaces~\cite{greeley2006computational}. Compositions based on Fe, Co,
	Mn, Ni, and Cu also rank near thermoneutrality in both models, including
	Fe$_{13}$Co$_3$ (rank~1), Ni$_7$Au (rank~3), Mn$_5$Ni$_7$ (rank~6),
	Cu$_2$Ni (rank~8), and Fe$_{13}$Cu (rank~15), all binary compositions
	absent from the Mamun training database. Of these, Fe$_{13}$Cu is also
	placed within the window by MatterSim (Section~\ref{sec:mlff_res}),
	whereas Ni$_7$Au and Mn$_5$Ni$_7$ are placed more strongly binding
	($-0.36$ and $-0.22$~eV).
	
	\begin{table}[h]
		\centering
		\caption{Representative top-ranked consensus HER candidates,
			ordered by consensus rank
			(mean $|\Delta G_{\mathrm{H}^*}|$ ascending).
			$\Delta G$ values (eV) are the best-site free-energy predictions of
			each model; ``sign'' marks agreement in the sign of $\Delta G$.
			Facet/site refer to the ALIGNN best site.
			$^\dagger$DFT candidates. Those outside the top~18 are appended at
			their consensus ranks. For the four candidates at ranks 62--123
			the DFT site differs from the best site listed here
			(Section~\ref{sec:dft_res}).}
		\label{tab:top20}
		\small
		\setlength{\tabcolsep}{3.5pt}
		\begin{tabular}{rlrrcl}
			\toprule
			Rank & Formula & $\Delta G_\mathrm{AL}$ & $\Delta G_\mathrm{MG}$
			& sign & Facet/site \\
			\midrule
			1 & Fe$_{13}$Co$_3$              & $+0.000$ & $-0.001$ & --         & (111) on-top \\
			2 & Pt$_{13}$Au$_3$              & $+0.001$ & $+0.001$ & \checkmark & (100) hollow \\
			3 & Ni$_7$Au                     & $+0.000$ & $+0.002$ & \checkmark & (100) bridge \\
			4 & In$_2$Pt$_{13}$              & $-0.001$ & $+0.001$ & --         & (111) hollow \\
			5 & IrPd$_7$                     & $+0.000$ & $+0.003$ & \checkmark & (110) hollow \\
			6 & Mn$_5$Ni$_7$                 & $+0.000$ & $+0.005$ & \checkmark & (100) hollow \\
			7 & Cu$_3$Pt$_7$                 & $+0.003$ & $-0.003$ & --         & (110) bridge \\
			8 & Cu$_2$Ni                     & $+0.000$ & $-0.010$ & --         & (100) bridge \\
			9 & AgPt$_4$                     & $+0.004$ & $+0.007$ & \checkmark & (110) hollow \\
			10 & Cu$_5$Ir                     & $-0.002$ & $+0.009$ & --         & (100) hollow \\
			11 & Pt$_7$Pb$^\dagger$           & $-0.000$ & $-0.013$ & \checkmark & (100) bridge \\
			12 & Pt$_4$Pb                     & $+0.012$ & $-0.001$ & --         & (110) bridge \\
			13 & Fe$_3$Ni$_{13}$              & $-0.001$ & $+0.013$ & --         & (110) bridge \\
			14 & Ni$_5$Au                     & $+0.001$ & $+0.015$ & \checkmark & (111) hollow \\
			15 & Fe$_{13}$Cu                  & $-0.006$ & $-0.013$ & \checkmark & (110) on-top \\
			16 & Cu$_2$Pt$_3^\dagger$         & $+0.003$ & $+0.018$ & \checkmark & (100) on-top \\
			17 & Co$_{11}$Ru$_5$              & $-0.013$ & $+0.011$ & --         & (100) bridge \\
			18 & Ga$_2$Ir                     & $-0.025$ & $+0.001$ & --         & (111) bridge \\
			20 & Cu$_5$Pt$_3^\dagger$         & $-0.000$ & $-0.030$ & \checkmark & (110) bridge \\
			40 & Pt$_5$Rh$^\dagger$           & $+0.109$ & $+0.001$ & \checkmark & (100) bridge \\
			62 & Pd$_{13}$Ru$_3^\dagger$      & $+0.070$ & $+0.119$ & \checkmark & (100) hollow \\
			76 & Cu$_3$Pd$_{13}^\dagger$      & $+0.053$ & $+0.185$ & \checkmark & (111) hollow \\
			102 & Pd$_5$Au$^\dagger$           & $+0.128$ & $+0.193$ & \checkmark & (111) hollow \\
			123 & TiPd$_5^\dagger$             & $+0.185$ & $+0.207$ & \checkmark & (111) bridge \\
			\bottomrule
		\end{tabular}
	\end{table}
	
	\subsection{MatterSim Energy-Difference Triangulation}
	\label{sec:mlff_res}
	
	Because the two GNN surrogates share the same Mamun training set, their
	agreement, however tight, could reflect a common bias rather than
	independent confirmation. We therefore added a third, independent
	estimate: MatterSim total-energy differences within the CHE
	framework (Eq.~\eqref{eq:CHE}), evaluated for the 25 lowest-$|\Delta G|$
	members of the 49-candidate high-confidence set, on the ALIGNN
	best-site configuration. MatterSim is a general-purpose force field
	trained on broad DFT data rather than on adsorption energies, so it
	carries a different systematic error from the BEEF-vdW-trained GNNs
	and serves as a triangulation before committing plane-wave DFT.
	
	The results are listed in Table~\ref{tab:mlff}. By construction these
	candidates lie within 0.12~eV of zero in both GNNs. MatterSim, which
	played no part in their selection, places 13 of the 25 within the window
	($|\Delta G_{\mathrm{H}^*}|\le 0.2$~eV) and nine of the remaining twelve
	on the strongly binding side. Its median offset from the mean GNN
	prediction is $-0.15$~eV, so for most candidates MatterSim binds hydrogen
	more strongly than the GNNs, the same direction in which DFT departs from
	all three surrogates (Section~\ref{sec:dft_res}).
	
	\begin{table}[h]
		\centering
		\caption{MatterSim energy-difference triangulation of the 25
			lowest-$|\Delta G|$ members of the 49-candidate high-confidence set.
			$\Delta G_\mathrm{AL}$ and $\Delta G_\mathrm{MG}$ are the ALIGNN and
			MEGNet best-site free-energy predictions; $\Delta G_\mathrm{MSim}$ is
			the independent MatterSim CHE value on the ALIGNN best-site
			configuration (Eq.~\eqref{eq:CHE}). All values in eV.
			$^\dagger$Computed with DFT.
			$^\ddagger$H incorporated into the top metal layer in the screened
			geometry (not carried forward).}
		\label{tab:mlff}
		\small
		\setlength{\tabcolsep}{6pt}
		\begin{tabular}{l rrr}
			\toprule
			Candidate & $\Delta G_\mathrm{AL}$ & $\Delta G_\mathrm{MG}$
			& $\Delta G_\mathrm{MSim}$ \\
			\midrule
			Pt$_{13}$Au$_3$              & $+0.001$ & $+0.001$ & $-0.226$ \\
			Ni$_7$Au                     & $+0.000$ & $+0.002$ & $-0.363$ \\
			IrPd$_7$                     & $+0.000$ & $+0.003$ & $-0.314$ \\
			Mn$_5$Ni$_7$                 & $+0.000$ & $+0.005$ & $-0.218$ \\
			AgPt$_4$                     & $+0.004$ & $+0.007$ & $-0.311$ \\
			Pt$_7$Pb$^\dagger$           & $-0.000$ & $-0.013$ & $-0.161$ \\
			Ni$_5$Au                     & $+0.001$ & $+0.015$ & $-0.267$ \\
			Fe$_{13}$Cu                  & $-0.006$ & $-0.013$ & $-0.070$ \\
			Cu$_2$Pt$_3^\dagger$         & $+0.003$ & $+0.018$ & $+0.085$ \\
			Cu$_5$Pt$_3^\dagger$         & $-0.000$ & $-0.030$ & $-0.056$ \\
			YCo$_5$                      & $-0.036$ & $-0.001$ & $+1.137$ \\
			Ni$_7$Pt                     & $-0.012$ & $-0.027$ & $-0.167$ \\
			Fe$_{11}$Co                  & $-0.053$ & $-0.007$ & $+0.048$ \\
			Ag$_5$Pt$_3^\ddagger$        & $+0.008$ & $+0.056$ & $+0.099$ \\
			ScPt$_{11}$                  & $+0.065$ & $+0.004$ & $-0.258$ \\
			Pd$_3$Pt$_{13}$              & $+0.065$ & $+0.006$ & $+0.274$ \\
			Fe$_7$Ni$_5$                 & $+0.038$ & $+0.040$ & $-0.200$ \\
			Ag$_7$Pt                     & $+0.072$ & $+0.010$ & $+0.207$ \\
			GaCo$_5$                     & $+0.018$ & $+0.072$ & $-0.192$ \\
			Cu$_5$Pt                     & $+0.018$ & $+0.073$ & $-0.059$ \\
			CuPd$_7$                     & $+0.022$ & $+0.070$ & $-0.227$ \\
			CoCu$_5$                     & $+0.088$ & $+0.009$ & $+0.179$ \\
			Pt$_5$Rh$^\dagger$           & $+0.109$ & $+0.001$ & $+0.139$ \\
			RuRh$_5$                     & $+0.003$ & $+0.113$ & $-0.080$ \\
			ZnNi$_4$                     & $+0.000$ & $+0.117$ & $-0.241$ \\
			\bottomrule
		\end{tabular}
	\end{table}
	
	The 13 candidates placed within the window by all three methods
	(Pt$_7$Pb, Fe$_{13}$Cu, Cu$_2$Pt$_3$, Cu$_5$Pt$_3$, Ni$_7$Pt,
	Fe$_{11}$Co, Ag$_5$Pt$_3$, Fe$_7$Ni$_5$, GaCo$_5$, Cu$_5$Pt, CoCu$_5$,
	Pt$_5$Rh, and RuRh$_5$) are the highest-confidence members of the
	shortlist. In this count MEGNet enters through its own best site. On
	the ALIGNN configuration evaluated by MatterSim, MEGNet also lies inside
	the window for five of the 13 (Pt$_7$Pb, Cu$_5$Pt$_3$, Fe$_{11}$Co,
	Ag$_5$Pt$_3$, and Pt$_5$Rh). Four of them (Pt$_7$Pb, Cu$_2$Pt$_3$, Cu$_5$Pt$_3$, and
	Pt$_5$Rh) were computed with DFT on these sites (Section~\ref{sec:dft_res}).
	Ag$_5$Pt$_3$ was not carried forward: in its screened geometry the H
	atom is fivefold coordinated and level with the top metal layer, that
	is, incorporated into the surface rather than adsorbed on it.
	
	\subsection{First-Principles DFT Validation}
	\label{sec:dft_res}
	
\begin{figure}[htbp]
		\centering
		\includegraphics[width=\linewidth]{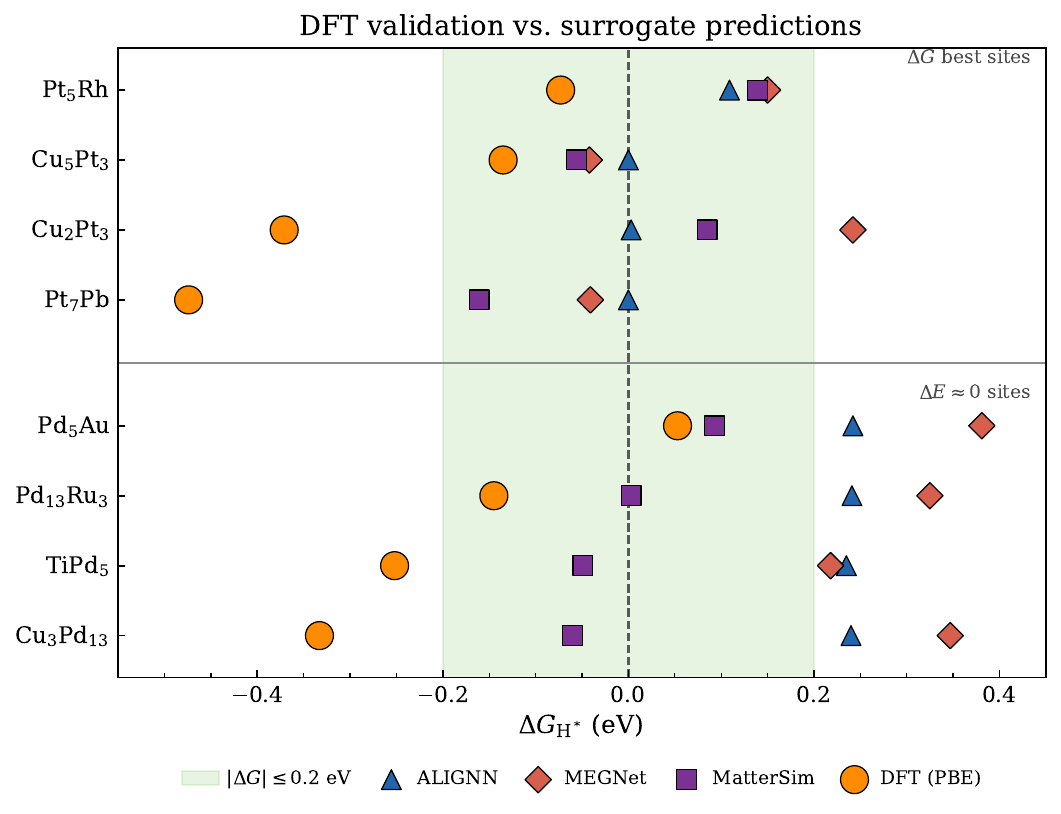}
		\caption{Hydrogen adsorption free energy $\Delta G_{\mathrm{H}^*}$
			for the eight DFT-validated candidates, grouped by site selection, comparing
			the plane-wave PBE DFT value with the ALIGNN and MEGNet predictions
			and the MatterSim energy-difference value on the same sites. The green
			band marks the thermoneutral window $|\Delta G|\le 0.2$~eV. By
			construction of the site selection, ALIGNN places the
			$\Delta E\approx0$ sites at $\Delta G\approx+0.24$~eV and the
			$\Delta G$ best sites close to zero. DFT binds hydrogen more strongly than all three
			surrogates for every candidate, and MatterSim lies closest to DFT.}
		\label{fig:dft}
	\end{figure}

	Eight sign-agreeing candidates of distinct chemistry were carried
	forward to DFT in two groups. Four of them, Pd$_5$Au, Pd$_{13}$Ru$_3$,
	TiPd$_5$, and Cu$_3$Pd$_{13}$, were selected for chemical diversity,
	well-defined (non-$P1$) bulk symmetry, and small cells rather than by
	consensus rank, and were computed on the ALIGNN sites closest to
	$\Delta E_{\mathrm{H}^*}=0$ (predicted
	$\Delta G_{\mathrm{H}^*}\approx+0.24$~eV). The other four, Pt$_7$Pb,
	Cu$_2$Pt$_3$, Cu$_5$Pt$_3$, and Pt$_5$Rh, are members of the 13
	candidates placed within the window by all three methods
	(Section~\ref{sec:mlff_res}) and were computed on their ALIGNN
	$\Delta G$ best sites. All eight DFT calculations therefore start from
	ALIGNN configurations, and the MEGNet values in Table~\ref{tab:dft} are
	its predictions on the same configurations. All eight are non-magnetic,
	with slab$+$H models of at most 49 atoms, and all except TiPd$_5$
	belong to the 49-member high-confidence list. TiPd$_5$ agrees in sign,
	but its MEGNet best site ($+0.21$~eV) lies just outside the window. The
	five earth-abundant alloys among the 13 (Fe$_{13}$Cu, Fe$_{11}$Co,
	Fe$_7$Ni$_5$, CoCu$_5$, and GaCo$_5$) contain magnetic elements and
	require spin-polarised calculations. Those attempted for Fe$_{13}$Cu
	and Fe$_7$Ni$_5$ did not converge within the available computational
	resources, so these alloys were left for future work. Table~\ref{tab:dft} and
	Fig.~\ref{fig:dft} compare the DFT results with the three surrogates.
	
	\begin{table*}[t]
		\centering
		\caption{DFT (PBE) validation of the eight selected candidates.
			$\Delta E_\mathrm{ads}$ is the electronic adsorption energy and
			$\Delta G_{\mathrm{H}^*}^{\mathrm{DFT}}=\Delta E_\mathrm{ads}+0.24$~eV
			(Eq.~\eqref{eq:CHE}). Site: H coordination in the DFT-relaxed geometry.
			$\Delta G_\mathrm{MSim}$, $\Delta G_\mathrm{AL}$, and $\Delta G_\mathrm{MG}$
			are the MatterSim, ALIGNN, and MEGNet values on the same site. Upper
			four rows: ALIGNN sites with $\Delta E_{\mathrm{H}^*}\approx0$. Lower
			four: ALIGNN $\Delta G$ best sites.}
		\label{tab:dft}
		\small
		\setlength{\tabcolsep}{5pt}
		\begin{tabular}{ll l rr rrr}
			\toprule
			Candidate & Facet & Site
			& $\Delta E_\mathrm{ads}$ & $\Delta G_{\mathrm{H}^*}^{\mathrm{DFT}}$
			& $\Delta G_\mathrm{MSim}$ & $\Delta G_\mathrm{AL}$ & $\Delta G_\mathrm{MG}$ \\
			&&& (eV) & (eV) & (eV) & (eV) & (eV) \\
			\midrule
			Pd$_5$Au         & (100) & hollow  & $-0.187$ & $+0.053$ & $+0.093$ & $+0.242$ & $+0.381$ \\
			Pd$_{13}$Ru$_3$  & (110) & bridge  & $-0.385$ & $-0.145$ & $+0.003$ & $+0.241$ & $+0.325$ \\
			TiPd$_5$         & (111) & bridge  & $-0.492$ & $-0.252$ & $-0.049$ & $+0.235$ & $+0.218$ \\
			Cu$_3$Pd$_{13}$  & (111) & hollow  & $-0.573$ & $-0.333$ & $-0.060$ & $+0.240$ & $+0.347$ \\
			\midrule
			Pt$_5$Rh         & (100) & hollow  & $-0.313$ & $-0.073$ & $+0.139$ & $+0.109$ & $+0.150$ \\
			Cu$_5$Pt$_3$     & (110) & hollow  & $-0.375$ & $-0.135$ & $-0.056$ & $-0.000$ & $-0.042$ \\
			Cu$_2$Pt$_3$     & (100) & on-top  & $-0.611$ & $-0.371$ & $+0.085$ & $+0.003$ & $+0.242$ \\
			Pt$_7$Pb         & (100) & bridge  & $-0.714$ & $-0.474$ & $-0.161$ & $-0.000$ & $-0.041$ \\
			\bottomrule
		\end{tabular}
	\end{table*}
	
	\begin{figure*}[t]
		\centering
		\includegraphics[width=0.94\textwidth]{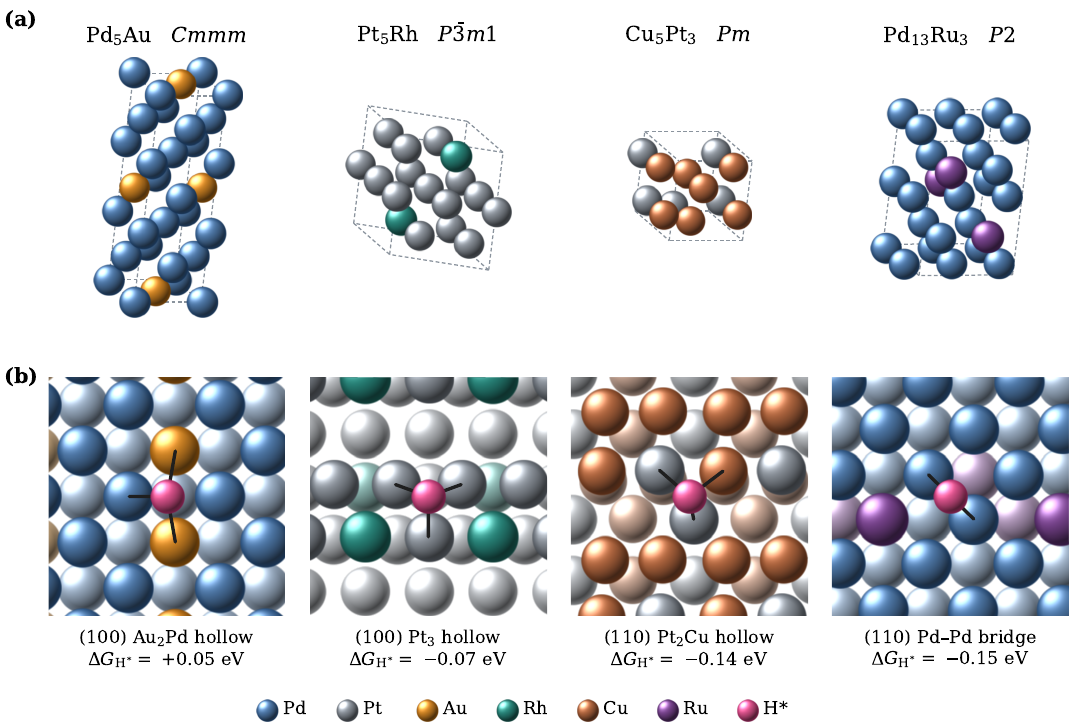}
		\caption{Structures of the four DFT-validated near-thermoneutral
			candidates. (a) Conventional bulk unit cells (dashed lines) of the
			MatterGen-generated alloys, labelled with formula and space group,
			drawn at a common scale. Pd$_5$Au, Pt$_5$Rh, and Pd$_{13}$Ru$_3$ are
			chemically ordered superstructures on an fcc parent lattice, and
			Cu$_5$Pt$_3$ has an hcp-like sublattice. (b) Top views of the
			DFT (PBE)-relaxed slab$+$H configurations, centred on the adsorbed
			H$^*$. Lines mark the H--metal bonds (within 2.2~\AA{}). Top-layer
			atoms and all atoms bonded to H are shown in full colour, and the
			deeper atoms in pale tones. Beneath each panel are the facet, the site
			named by the metal atoms bonded to H, and the DFT
			$\Delta G_{\mathrm{H}^*}$.}
		\label{fig:structures}
	\end{figure*}
	
	All DFT relaxations converged. DFT places four of the eight candidates
	inside the thermoneutral window, Pd$_5$Au, Pt$_5$Rh, Cu$_5$Pt$_3$, and
	Pd$_{13}$Ru$_3$, and the other four bind hydrogen more strongly.
	Pd$_5$Au ($+0.05$~eV) and Pt$_5$Rh ($-0.07$~eV) are the closest to
	thermoneutral. The four in-window alloys, shown in
	Fig.~\ref{fig:structures}, are the prime targets for synthesis and
	electrochemical testing. They are rich in platinum-group metals, which
	reflects the exclusion of the magnetic earth-abundant alloys from DFT
	rather than a preference of the screen. Cu$_5$Pt$_3$ shows that Pt can
	be strongly diluted while hydrogen binding stays near thermoneutral.
	
	All three surrogates underbind hydrogen relative to DFT for every
	candidate. MatterSim deviates least (mean absolute deviation 0.22~eV,
	against 0.35~eV for ALIGNN and 0.41~eV for MEGNet) and best reproduces
	the DFT ordering (Spearman rank correlation 0.67, against 0.52 and
	0.19). Two effects make the GNN deviations larger than the test-set MAE
	(0.16--0.17~eV). First, choosing the site closest to zero among some
	50 predictions per bulk favours sites whose prediction errors carry
	them toward zero. Second, the GNN labels are BEEF-vdW energies, and on
	close-packed Pt, Pd, Rh, and Ni surfaces PBE binds hydrogen more
	strongly than BEEF-vdW by 0.12--0.23~eV per H atom relative to
	$\tfrac12\mathrm{H_2}$, while lying closer to the measured values on
	four of five benchmark surfaces~\cite{wellendorff2015benchmark,duanmu2017validation}.
	MatterSim, fitted to PBE-level data, shows an offset of the same sign,
	so the functional cannot account for all of it. Because of this offset,
	two of the in-window alloys, Pd$_5$Au and Pd$_{13}$Ru$_3$, come from
	the $\Delta E\approx0$ group, which ALIGNN places just outside the
	window on the weak-binding side, while two $\Delta G$ best sites
	predicted near zero, Cu$_2$Pt$_3$ and Pt$_7$Pb, fall below the window in
	DFT. Seven of the eight bulks are fcc-derived, but six of the eight DFT
	surfaces are more open than the close-packed Mamun terminations (two to
	four in-plane neighbours per surface atom, against six), so the
	extrapolation from the training data lies mainly in the surface
	terminations and chemical orderings rather than in the parent lattice.
	Within each group ALIGNN places the candidates within about 0.1~eV of
	one another by construction, and MEGNet does not resolve them either.
	This is the gap that the MatterSim triangulation and the DFT validation
	were introduced to close.
	
	In summary, half of the DFT-tested candidates bind hydrogen within the
	thermoneutral window, and all four are absent from the Mamun training
	set. Pd$_{13}$Ru$_3$ is also absent from Alex-MP, whereas Pd$_5$Au,
	Pt$_5$Rh, and Cu$_5$Pt$_3$ match entries of that reference set, so for
	these three the new result is their HER activity rather than the
	structure. The independent MatterSim estimate is the more reliable
	guide to the DFT value and ordering.
	
	\subsection{Dynamical Stability of the DFT Candidates}
	\label{sec:phonon_res}
	
\begin{figure*}[p]
		\centering
		\includegraphics[width=0.86\textwidth]{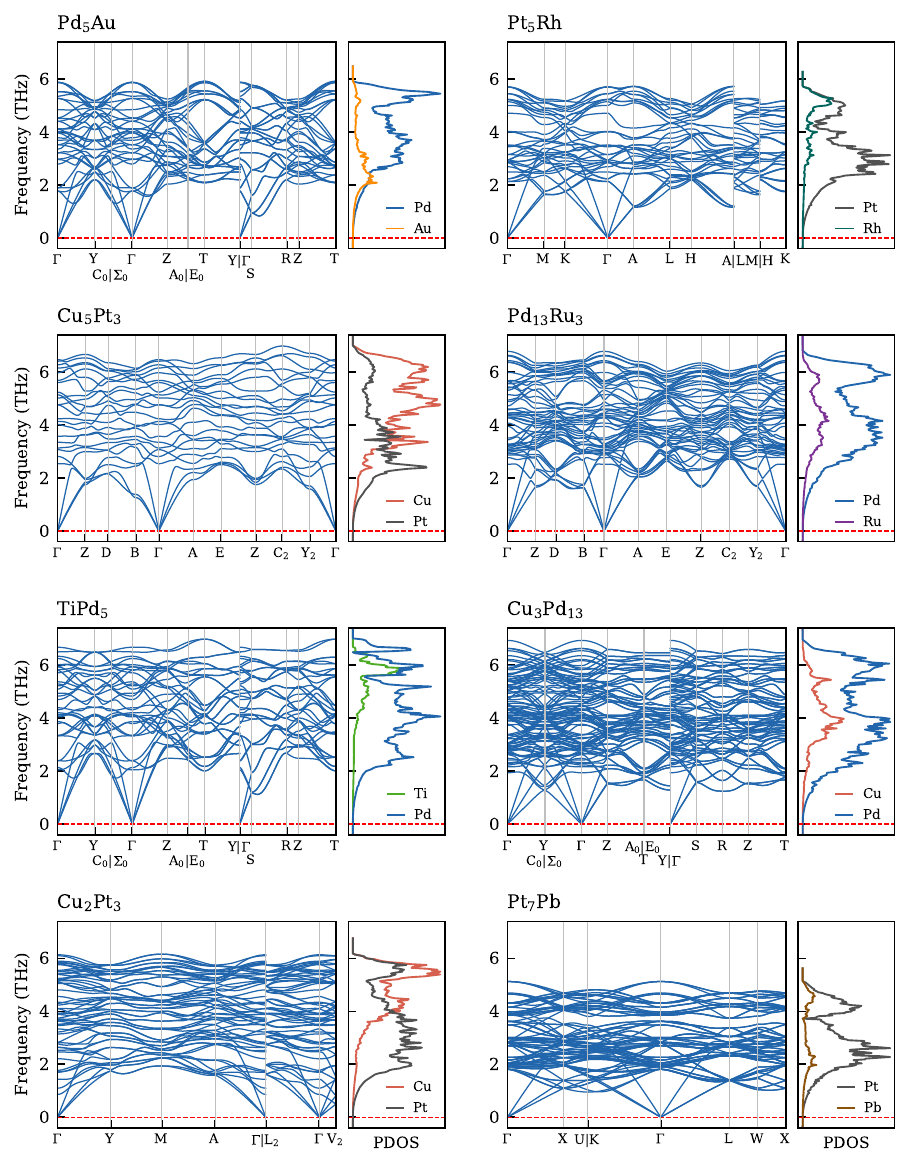}
		\caption{Phonon dispersion and element-projected phonon density of
			states (PDOS) for the eight DFT-validated candidates, ordered by DFT
			$\Delta G_{\mathrm{H}^*}$ from weakest to strongest hydrogen binding
			(left to right, top to bottom), computed with MatterSim forces and the
			finite-displacement method.}
		\label{fig:phonon}
	\end{figure*}

	All eight DFT candidates are dynamically stable. Their phonon
	dispersions (Fig.~\ref{fig:phonon}), computed with MatterSim forces
	(Section~\ref{sec:phonon_meth}), show no imaginary modes beyond
	numerical noise below 0.01~THz in magnitude, so each bulk structure sits
	at a local minimum of the potential energy surface, a necessary
	condition for synthesisability.
	
	\section{Conclusions}
	\label{sec:conc}
	
	We have presented a computational pipeline for HER catalyst discovery
	that combines a fine-tuned diffusion-based generative model (MatterGen),
	a machine-learning force field for surface screening and slab
	relaxation (MatterSim), and two GNN surrogates (MEGNet and ALIGNN) for
	hydrogen adsorption. Fine-tuned on 1{,}569 bulk alloys labelled with
	BEEF-vdW hydrogen adsorption energies, the generator produced 3{,}000
	structures, 1{,}151 of them stable and unique, from which 730 binary
	alloys of the training-set metals absent from the Mamun database were
	screened.
	
	On a common test set of 1{,}330 slab$+$H structures the two surrogates
	reach comparable accuracy (MAE 0.161~eV for ALIGNN and 0.174~eV for
	MEGNet) through different inductive biases, pairwise distances in
	MEGNet and three-body bond angles in ALIGNN. Applied to 39{,}132
	slab$+$H configurations on the three most stable surfaces of each bulk,
	they agree well (Pearson $r=0.847$) and place 77 bulk alloys within
	$\pm0.2$~eV in both models, of which 49 sign-agreeing and converged
	alloys form a high-confidence set.
	
	MatterSim energy differences place 13 of the 25 lowest-$|\Delta G|$
	high-confidence candidates within the window, and DFT (PBE) on eight
	candidates places four, Pd$_5$Au, Pt$_5$Rh, Cu$_5$Pt$_3$, and
	Pd$_{13}$Ru$_3$, within the thermoneutral window. All eight are
	dynamically stable in MatterSim phonon calculations. All three
	surrogates underbind hydrogen relative to DFT. MatterSim deviates least
	(mean absolute deviation 0.22~eV) and tracks the DFT ordering best,
	whereas the GNNs deviate by 0.35--0.41~eV, so an independent third
	method is essential.
	
	The four in-window alloys are priority targets for synthesis and
	electrochemical testing. Some Pt--Rh \cite{Zhang2025} and Cu--Pt compounds \cite{Kaya2021, Shen2024}, although not at the stoichiometries proposed here, have recently been synthesized and tested as HER electrocatalysts. The reported performances, comparable to or higher than those of Pt-based references, support that the present pipeline can be efficient at targeting promising systems for HER catalysis. The other eight alloys placed within the window by all three methods, in particular the earth-abundant Fe-, Co-, and Ni-based ones, and more broadly the 49 high-confidence alloys, form a prioritised list for further DFT validation. The pipeline can be extended to the oxygen evolution reaction (OER) and nitrogen reduction reaction (NRR) by retraining the surrogates on the corresponding adsorption-energy databases.
	
	\section*{Data and Code Availability}
	
	The Mamun training dataset is publicly available at
	Catalysis-Hub.org (\url{https://www.catalysis-hub.org})~\cite{mamun2019high}.
	MatterGen source code is at
	\url{https://github.com/microsoft/mattergen}~\cite{zeni2025generative}.
	The MEGNet and ALIGNN implementations used here are the
	open-source Materials Graph Library (MatGL)~\cite{ko2025matgl} and
	ALIGNN~\cite{choudhary2021atomistic} packages.
	The code and data for MatterGen fine-tuning and generation, slab and
	adsorption-site construction, MatterSim relaxation, GNN training and
	screening, and consensus analysis, together with the surface-energy
	screen and all 39{,}132 relaxed slab$+$H structures, are available at
	\url{https://github.com/Vipinkurian/HER_gen}. The DFT input files,
	the phonon calculations, and any further data are available from the
	corresponding author upon reasonable request.
	
	\section*{Author Contributions}
	V.K.E.\ designed and implemented the computational pipeline,
	carried out the MatterGen fine-tuning and generation, slab
	construction, MatterSim relaxations, GNN training and screening,
	and the DFT calculations, performed the data analysis,
	and wrote the original draft. S.L.T.\ contributed to the interpretation of the alloy chemistry and first-principles results, acquired funding, and reviewed and edited the manuscript. H.V.\ conceived and
	supervised the study, acquired funding, and reviewed and edited
	the manuscript. All authors discussed the results and approved the
	final manuscript.
	
	\section*{Competing Interests}
	The authors declare no competing interests.
	
	\section*{Acknowledgements}
	This work was supported by the University of Caen from the CAI4SCIENCE project. Part of this work was performed using computing resources of CRIANN (Normandy, France).
	

	\clearpage
	\section*{Supplementary Information}
	\setcounter{figure}{0}
	\renewcommand{\thefigure}{S\arabic{figure}}
	\renewcommand{\theHfigure}{S\arabic{figure}}
	
	\begin{figure}[h]
		\centering
		\includegraphics[width=\linewidth]{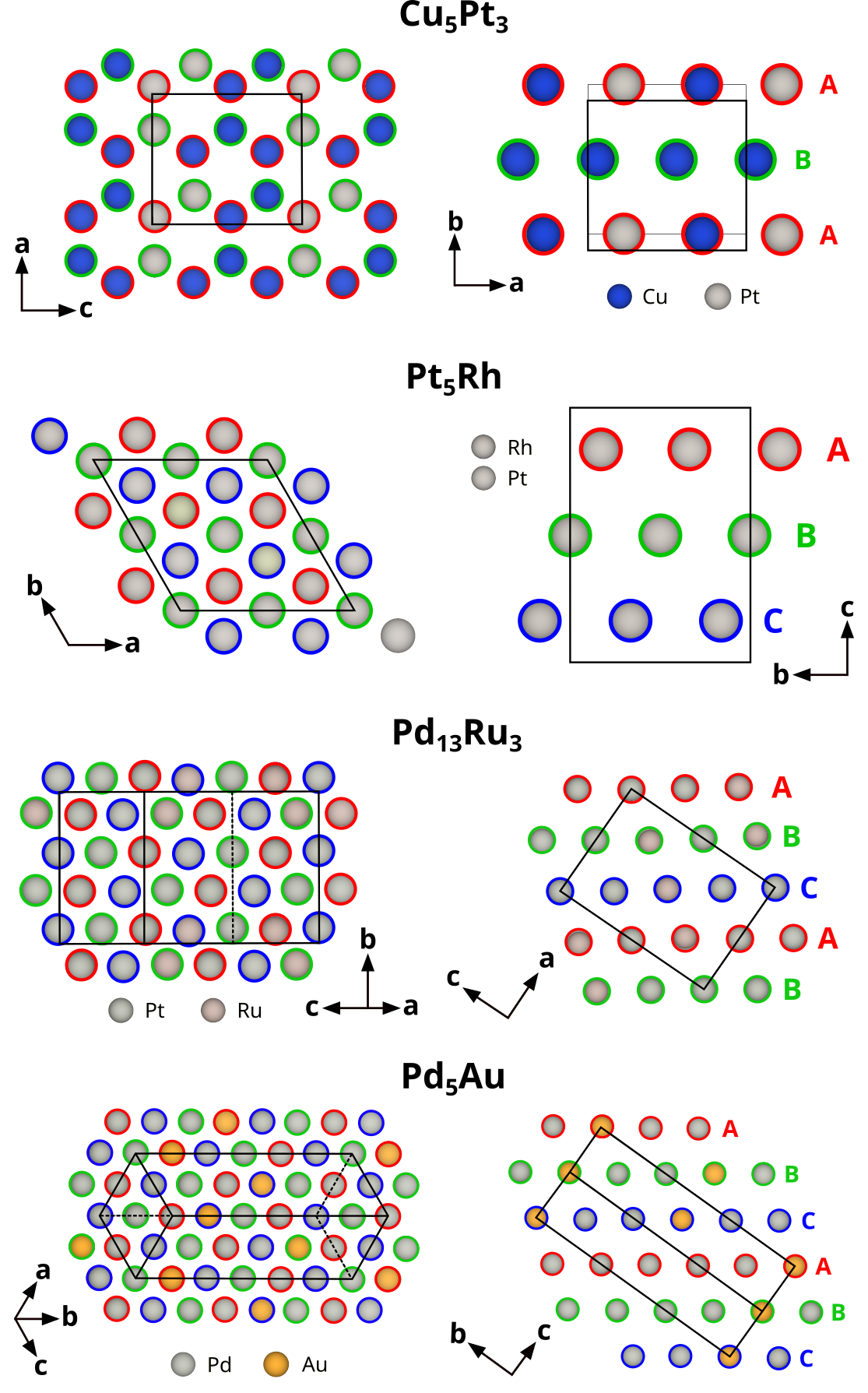}
		\caption{Stacking of the close-packed metal layers in the four DFT
			candidates that bind hydrogen within the thermoneutral window.}
		\label{fig:S1}
	\end{figure}
	
\end{document}